CAN A PLANTAR PRESSURE-BASED TONGUE-PLACED ELECTROTACTILE BIOFEEDBACK

IMPROVE POSTURAL CONTROL UNDER ALTERED VESTIBULAR AND NECK PROPRIOCEPTIVE

CONDITIONS?

Nicolas VUILLERME[1], Olivier CHENU[1], Nicolas PINSAULT[1], Anthony FLEURY[1],

Jacques DEMONGEOT[1] and Yohan PAYAN[1]

[1] Laboratoire TIMC-IMAG, UMR UJF CNRS 5525, La Tronche, France

**Section editor**: Dr. David Fitzpatrick

**Address for correspondence:**

Nicolas VUILLERME

Laboratoire TIMC-IMAG, UMR UJF CNRS 5525

Faculté de Médecine

38706 La Tronche cédex

France.

Tel: (33) (0) 4 76 63 74 86

Fax: (33) (0) 4 76 51 86 67

Email: nicolas.vuillerme@imag.fr

**ABBREVIATIONS**

TDU: Tongue Display Unit

CoP: center of foot pressure

CNS: central nervous system

DZ: Dead Zone

ANOVA: analysis of variance

SD: standard deviation

**ABSTRACT**

We investigated the effects of a plantar pressure-based tongue-placed electrotactile biofeedback on postural control during quiet standing under normal and altered vestibular and neck proprioceptive conditions. To achieve this goal, fourteen young healthy adults were asked to stand upright as immobile as possible with their eyes closed in two Neutral and Extended head postures and two conditions of No-biofeedback and Biofeedback. The underlying principle of the biofeedback consisted of providing supplementary information related to foot sole pressure distribution through a wireless embedded tongue-placed tactile output device. Centre of foot pressure (CoP) displacements were recorded using a plantar pressure data acquisition system. Results showed that (1) the Extended head posture yielded increased CoP displacements relative to the Neutral head posture in the No-biofeedback condition, with a greater effect along the anteroposterior than mediolateral axis, whereas (2) no significant difference between the two Neutral and Extended head postures was observed in the Biofeedback condition. The present findings suggested that the availability of the plantar pressure-based tongue-placed electrotactile biofeedback allowed the subjects to suppress the destabilizing effect induced by the disruption of vestibular and neck proprioceptive inputs associated with the head extended posture. These results are discussed according to the sensory re-weighting hypothesis, whereby the central nervous system would dynamically and selectively adjust the relative contributions of sensory inputs (i.e., the sensory weights) to maintain upright stance depending on the sensory contexts and the neuromuscular constraints acting on the subject.

**INTRODUCTION**

Biofeedback systems for balance control consist in supplying individuals with additional artificial information about body orientation and motion to supplement the natural visual, somatosensory and vestibular sensory cues. Considering the important contribution of plantar cutaneous information in the regulation of postural sway during quiet standing (e.g. Kavounoudias et al., 1998; Meyer et al., 2004; Vuillerme and Pinsault, 2007), we recently developed a biofeedback system whose underlying principle consists in supplying the user with supplementary sensory information related to foot sole pressure distribution through a tongue-placed tactile output device generating electrotactile stimulation of the tongue (Vuillerme et al., 2007b,c,d,e, 2008). There were several reasons to use the tongue as a substrate for electrotactile stimulation (Bach-y-Rita et al., 1998, 2003). Because of its dense mechanoreceptive innervations (Trulsson and Essick, 1997) and large somatosensory cortical representation (Picard and Olivier, 1983), the tongue can convey higher-resolution information than the skin can (Sampaio et al., 2001; van Boven and Johnson, 1994). In addition, due to the excellent conductivity offered by the saliva, electrotactile stimulation of the tongue can be applied with much lower voltage and current than is required for the skin (Bach-y-Rita et al., 1998). Finally, the tongue is in the protected environment of the mouth and is normally out of sight and out of the way, which could make a tongue-placed tactile display aesthetically acceptable.

While the effectiveness of this biofeedback in improving postural control during quiet standing has recently been demonstrated in young healthy subjects, under reliable sensory conditions and normal neuromuscular state (Vuillerme et al., 2007b,c,d,e, 2008), whether the central nervous system (CNS) is able to integrate this biofeedback when subjected to challenging postural conditions is yet to be established. In the present experiment, we assessed the postural effects of a plantar pressure-based tongue-placed electrotactile biofeedback under normal and altered conditions of vestibular and neck proprioceptive sensory information.

It was hypothesised that:

(1) the alteration of vestibular and neck proprioceptive information would increase centre of foot pressure (CoP) displacements,

(2) the availability of plantar pressure-based tongue-placed electrotactile biofeedback would decrease CoP displacements, and

(3) the availability of plantar pressure-based tongue-placed electrotactile biofeedback would limit the destabilizing effect induced by the alteration of vestibular and neck proprioceptive information.

**EXPERIMENTAL PROCEDURES**

<u>Subjects</u>

Fourteen young healthy university students (age: $25.0 \pm 3.8$ years; body weight: $69.9 \pm 11.8$ kg; height: $175.0 \pm 10.5$ cm; mean $\pm$ SD) participated in the experiment. Subjects had to be healthy without a history of neck pain, neurological or vestibular impairment, injury or operation in the cervical spine. They gave their informed consent to the experimental procedure as required by the Helsinki declaration (1964) and the local Ethics Committee.

<u>Task and procedure</u>

Eyes closed, subjects stood barefoot, feet together, their hands hanging at the sides, on a plantar pressure data acquisition system (Force Sensitive Applications (FSA) Orthotest Mat, Vista Medical Ltd.). Subject's task was to sway as little as possible in two Neutral and Extended head postures. The head extended posture is recognised to induce (1) a modification in the orientation of the vestibular organs that may place the utricular otoliths well beyond their working range and render balance related vestibular information unreliable to the CNS (e.g., Brandt et al., 1981, 1986; Jackson and Epstein, 1991; Straube et al., 1992) and (2) abnormal sensory inputs arising from neck proprioceptors (e.g. Jackson and Epstein, 1991; Karlberg, 1995; Ryan and Cope, 1955), that represent a challenge for the postural control system. In the Neutral head posture, subjects were asked to keep their head in a straight-ahead direction. In the Extended head posture, they were asked to tilt their head backward for at least 45° in the sagittal plane as previously done by other authors (e.g., Anand et al., 2002, 2003; Brandt et al., 1981; Buckley et al., 2005; Jackson and Epstein, 1991; Simoneau et al., 1992; Vuillerme and Rougier, 2005). The experimenter always stood by the subjects to monitor their posture and their head position throughout the trial. Subjects were asked to adopt the required posture and to stabilise their body sway for a 40 s period. These two postures were executed in two conditions of No-biofeedback and Biofeedback. Regardless of the experimental condition, the first 10 s of each trial were not considered for further analyses. In the Biofeedback condition, this 10 s period was used to scale the threshold of a plantar pressure-based, tongue-placed tactile biofeedback system with which subjects were asked to perform the postural task. This biofeedback system comprises two major components: (1) the sensory unit and (2) the tongue-placed tactile output unit. The plantar pressure data acquisition system (Force Sensitive Applications (FSA) Orthotest Mat, Vista Medical Ltd.) was used as sensory unit. This pressure mat (sensing area : $350 \times 350$ mm = $122500$ mm²), contains a $32 \times 32$ grid of piezo resistive sensors (sensor number : 1024; sensors dimensions : $3.94 \times 3.94$ mm ; space between sensors : 2.7 mm ; 0.84 sensor/cm²), allowing the magnitude of pressure exerted on each left and right foot sole at each sensor location to be transduced into the calculation of the positions of the resultant CoP (sampling frequency : 10 Hz). Resultant CoP data were then fed back in real time to a tongue-placed tactile output device (temporal latency : 300 ms). This so-called Tongue Display Unit (TDU), initially introduced by Bach-y-Rita et al. (1998, 2003), comprises a 2D array (15 × 15 mm) of 36 electrotactile electrodes each with a 1.4 mm diameter, arranged in a 6 × 6 matrix positioned in close contact with the anterior-superior surface of the tongue. While this matrix of electrodes was originally connected to an external electronic device via a flat cable passing out of the mouth in a previous version (e.g. Vuillerme et al., 2007a,b,c,e), we recently developed a wireless radio-controlled version of this tongue-placed tactile output device (Vuillerme et al., 2007d, 2008) including microelectronics, antenna and battery, which can be worn inside the mouth like an orthodontic retainer (Figure 1). In the Biofeedback condition, subjects were asked to actively and carefully hold their tongue against the matrix of electrodes.

------------------------------------

Please insert Figure 1 about here

------------------------------------

The underlying principle of this biofeedback system was to supply the user with supplementary information about the position of the CoP relative to a predetermined adjustable "dead zone" (DZ) through the TDU (Vuillerme et al., 2007b,c,d,e) (Figure 2).

------------------------------------

Please insert Figure 2 about here

------------------------------------

As mentioned above, in the present experiment, anteroposterior and mediolateral bounds of the DZ were set as the standard deviation of subject's CoP displacements recorded for 10 s preceding each experimental trial.

To avoid an overload of sensory information presented to the user, a simple and intuitive coding scheme for the TDU, consisting in a "threshold-alarm" type of feedback rather than a continuous feedback about ongoing position of the CoP, was then used:

(1) when the position of the CoP was determined to be within the DZ, no electrical activation of any of the electrodes of the matrix was provided;

(2) for the entire time the position of the CoP was determined to be outside the DZ, - i.e., when it was most needed, electrical activation of either the anterior, posterior, right or left zone of the matrix (1 × 4 electrodes) (i.e. electrotactile stimulation of front, rear, right of left portion of the tongue) was provided depending on whether the actual position of the CoP was in a too anterior, posterior, right or left position relative to the DZ, respectively. Interestingly, this type of sensory coding scheme for the TDU allows the activation of distinct and exclusive electrodes for a given position of the CoP with respect to the DZ (Figure 2).

Finally, in the present experiment, the intensity of the electrical stimulating current was adjusted for each subject, and for each of the front, rear, right and left portions of the tongue.

Several practice runs were performed prior to the test to ensure that subjects had mastered the relationship between the position of the CoP relative to the DZ and lingual stimulations.

Five trials for each condition were recorded. The order of presentation of the two Neutral and Extended head postures and the No-biofeedback and Biofeedback conditions was counterbalanced. Subjects were not given feedback about their postural performance.

Analysis

Two dependent variables were used to describe subject's postural behaviour: (1) the standard deviation and (2) the range extracted from ~~of~~ the CoP displacements along the mediolateral and anteroposterior axes averaged for the last 30 s period of each trial. The calculation of the standard deviation of the CoP displacements provides a measure of spatial variability of CoP around the mean position. The range of the CoP displacements indicates the difference between the maximum and minimum values of the CoP. A large value in the range of the CoP displacements indicates that the resultant forces are displaced towards the balance stability boundaries of the participant and could challenge their postural stability (e.g. Patton et al. 2000).

Two Head postures (Neutral *vs.* Extended) × 2 Biofeedback (No-biofeedback *vs.* Biofeedback) × 2 Axes (Mediolateral *vs.* Anteroposterior) analyses of variances (ANOVAs) with repeated measures on all factors were applied to the data. Post-hoc analyses (*Newman-Keuls*) were performed whenever necessary. Level of significance was set at 0.05.

**RESULTS**

Analysis of the standard deviation of the CoP displacements showed main effects of Head posture ($F(1,13) = 13.47$, $P < 0.01$) and Biofeedback ($F(1,13) = 11.22$, $P < 0.01$), two significant two way-interactions of Head posture × Biofeedback ($F(1,13) = 8.38$, $P < 0.05$) and Head posture × Axis ($F(1,13) = 6.87$, $P < 0.05$), and a significant three-way interaction of Head posture × Biofeedback × Axis ($F(1,13) = 10.73$, $P < 0.01$). As illustrated in Figure **3**, the decomposition of the three-way interaction into its simple main effects indicated that (1) the Extended head posture yielded larger standard deviation of the CoP displacements relative to the Neutral head position in the No-biofeedback condition ($Ps < 0.01$), (2) this destabilizing effect was more accentuated along the anteroposterior ($P < 0.001$) than mediolateral axis ($P < 0.01$), whereas (3) no significant difference between the two Neutral and Extended head postures was observed in the Biofeedback condition ($Ps > 0.05$).

------------------------------------

Please insert Figure 3 about here

------------------------------------

Results obtained for the range of the CoP displacements were consistent with those obtained for the standard deviation of the CoP displacements. The ANOVA showed main effects of Head posture ($F(1,13) = 16.25$, $P < 0.01$) and Biofeedback ($F(1,13) = 11.60$, $P < 0.01$), two significant two way-interactions of Head posture × Biofeedback ($F(1,13) = 8.43$, $P < 0.05$) and Head posture × Axis ($F(1,13) = 5.32$, $P < 0.05$), and a significant three-way interaction of Head posture × Biofeedback × Axis ($F(1,13) = 9.92$, $P < 0.01$). As illustrated in Figure 4, the decomposition of the three-way interaction into its simple main effects indicated that (1) the Extended head posture yielded larger range of the CoP displacements relative to the Neutral head posture in the No-biofeedback condition ($Ps < 0.01$), (2) this destabilizing effect was more accentuated along the anteroposterior ($P < 0.001$) than mediolateral axis ($P < 0.001$), whereas (3) no significant difference between the two Neutral and Extended head postures was observed in the Biofeedback condition ($Ps > 0.05$).

------------------------------------

Please insert Figure 4 about here

------------------------------------

## DISCUSSION

We investigated the effects of a plantar pressure-based tongue-placed electrotactile biofeedback on postural control during quiet standing under normal and altered vestibular and neck proprioceptive conditions.

To achieve this goal, fourteen young healthy adults were asked to stand upright as immobile as possible with their eyes closed in two Neutral and Extended head postures and two conditions of No-biofeedback and Biofeedback. The underlying principle of the biofeedback consisted of providing supplementary information related to foot sole pressure distribution through a wireless embedded tongue-placed tactile output device (Figure 1 and Figure 2).

On the one hand, these results showed that the extended head posture deteriorated postural control, with a greater destabilizing effect along the anteroposterior than mediolateral axis, as indicated by the significant interactions of Head posture × Axis observed for the standard deviation (Figure 3) and range of the CoP displacements (Figure 4). These results were expected (*hypothesis 1*), in accordance with previous observations (e.g., Anand et al., 2002, 2003; Brandt et al., 1981, 1986; Buckley et al., 2005; Jackson and Epstein, 1991; Kogler et al., 2000; Norré, 1995; Paloski et al., 2006; Simoneau et al., 1992; Straube et al., 1992; Vuillerme and Rougier, 2005).

On the other hand, the availability of the biofeedback improved postural control, as indicated by the decreased standard deviation (Figure 3) and range of the CoP displacements (Figure 4) observed in the Biofeedback relative to the No-biofeedback condition. This result also was expected (*hypothesis 2*). It confirms the ability of the CNS to efficiently integrate an artificial plantar pressure information delivered through electrotactile stimulation of the tongue to improve postural control during quiet standing (Vuillerme et al., 2007b,c,d,e, 2008).

Finally, the availability of the biofeedback allowed the subjects to suppress the destabilizing effect induced by the extension of the head, as indicated by the significant interactions Head posture × Biofeedback observed for the standard deviation (Figure 3) and the range the CoP displacements (Figure 4). This result confirms our *hypothesis 3*. Furthermore, as indicated by the significant three-way interactions of Head posture × Biofeedback × Axis observed for the standard deviation (Figure 3) and the range of the CoP displacements (Figure 4), the stabilizing effect of the Biofeedback was more pronounced along the anteroposterior than mediolateral axis. Together with the greater destabilizing effect of the extended head posture observed along the anteroposterior than the mediolateral axis, these results suggest that the effectiveness of the biofeedback in reducing the CoP displacements depends on the amount of postural sway observed when the biofeedback was not available. Interestingly, this finding is consistent with a recent study, reporting that the degree of postural stabilization induced by the use of a plantar pressure-based tongue-placed electrotactile depends on subject's balance control capabilities, the biofeedback yielding a greater stabilizing effect in subjects exhibiting the greatest CoP displacements when standing in the No-biofeedback condition (Vuillerme et al., 2007c). At this point, another possible reason leading to these results could be that the use of the tongue-placed electro-tactile biofeedback may have lead subjects to pay more attention to the regulation of their CoP displacements. However, in a recent study, in which were instructed to deliberately focus their attention on their body sway and to increase their active intervention into postural control, postural oscillations were not reduced (Vuillerme and Nafati, 2007). We thus believe that the postural improvement observed in the Biofeedback condition could not be attributed to the subjects' paying more attention to the regulation of their CoP displacements, but rather to their ability to effectively integrate the artificial plantar pressure information delivered through electro-tactile stimulation of the tongue.

On the whole, the present results could be attributable to sensory re-weighting hypothesis (e.g. Horak and MacPherson, 1996; Peterka, 2002; Peterka and Loughlin, 2004; Vuillerme et al., 2001, 2002, 2005, 2006a; Vuillerme and Pinsault, 2007), whereby the CNS dynamically and selectively adjusts the relative contributions of sensory inputs (i.e., the sensory weights) to maintain upright stance depending on the sensory contexts and the neuromuscular constraints acting on the subject. An example of this is the adaptive capabilities of the postural control system to cope with a degradation of proprioceptive signals from the ankle consecutive to muscle fatigue. In condition of ankle muscle fatigue, indeed, the sensory integration process has been shown to (1) decrease the contribution of proprioceptive cues from the ankle, degraded by the fatiguing exercise (Vuillerme et al., 2007a), and (2) increase the contribution of vision (Ledin et al., 2004; Vuillerme et al., 2006a), cutaneous inputs from the foot and shank (Vuillerme and Demetz, 2007) and haptic cues from the finger (Vuillerme and Nougier, 2003), providing reliable and accurate sensory information for controlling posture. Following this train of thought, the decreased CoP displacements observed in the Extended head posture when the Biofeedback was in use relative to when it was not suggests an increased reliance on sensory information related to the plantar pressure, in condition of altered vestibular and neck proprioceptive information. Interestingly, this interpretation is consistent with the increased postural responses to the alteration of somatosensory information from the foot and ankle, either induced by requiring healthy subjects to stand on a compliant (Anand et al., 2002, 2003; Buckley et al., 2005) or on a sway-referenced (Kogler et al., 2000; Paloski et al., 2006) support surface, or by applying vibratory proprioceptive stimulation to the calf muscles (Ledin et al., 2003), previously observed in an extended relative to a neutral head posture.

Finally, in addition to its relevance on the field of neuroscience, the present findings also could have implications in clinical and rehabilitative areas for restoring balance control in individuals with impaired vestibular and/or neck proprioceptive capacity. Investigations involving patients with chronic whiplash injury and with vestibular loss are planned to address this issue.

**Acknowledgements**

The authors are indebted to Professor Paul Bach-y-Rita for introducing us to the Tongue Display Unit and for discussions about sensory substitution. Paul has been for us more than a partner or a supervisor: he was a master inspiring numerous new fields of research in many domains of neurosciences, biomedical engineering and physical rehabilitation. This research was supported by the company IDS and the Fondation Garches. The company Vista Medical is acknowledged for supplying the FSA pressure mapping system. The authors would like to thank the anonymous reviewers for helpful comments and suggestions. Special thanks also are extended to Claire H. for various contributions.

**Figure captions**

**Figure 1.** Photograph of the wireless radio-controlled tongue-placed tactile output device used in the present experiment. It consists in a 2D electrodes array arranged in a $6 \times 6$ matrix glued onto the inferior part of the orthodontic retainer which also includes microelectronics, antenna and battery.

**Figure 2.** Principle of the plantar pressure-based tongue-placed electrotactile biofeedback for balance.

The central black rectangle located on the base of support and the thin black trace inside represent the predetermined dead zone (DZ) and the trajectory of the CoP (central panel).

The four grey squares and the black dots inside represent the $6 \times 6$ matrix of electrotactile electrodes of the wireless radio-controlled version of the Tongue Display Unit (TDU) maintained in contact with the anterior-superior surface of the tongue, and the activated electrodes, respectively.

There were 5 possible stimulation patterns of the TDU:

**(1)** no electrical activation of any of the electrodes of the matrix was provided when the CoP position was determined to be within the DZ (*central panel*).

**(2)** electrical activation of either the anterior, posterior, right or left zone of the matrix ($1 \times 4$ electrodes) were provided, when the CoP positions were determined to be outside the DZ, located towards the front, rear, left and right of the DZ, respectively (four *peripheral panels*). These 4 stimulation patterns correspond to the stimulations of the front, rear, left and right portions of the tongue dorsum, respectively."

**Figure 3.** Mean and standard deviation of the standard deviation of the CoP displacements along the mediolateral and anteroposterior directions obtained in the two *Neutral* and *Extended* head postures and the two conditions of No-biofeedback and Biofeedback. The two conditions of No-biofeedback and Biofeedback are presented with different symbols: No-biofeedback (*white circle*) and Biofeedback (*black square*).

**Figure 4.** Mean and standard deviation of the range of the CoP displacements along the mediolateral and anteroposterior directions obtained in the two *Neutral* and *Extended* head postures and the two conditions of No-biofeedback and Biofeedback. The two conditions of No-biofeedback and Biofeedback are presented with different symbols: No-biofeedback (*white circle*) and Biofeedback (*black square*).

**Figure 1**

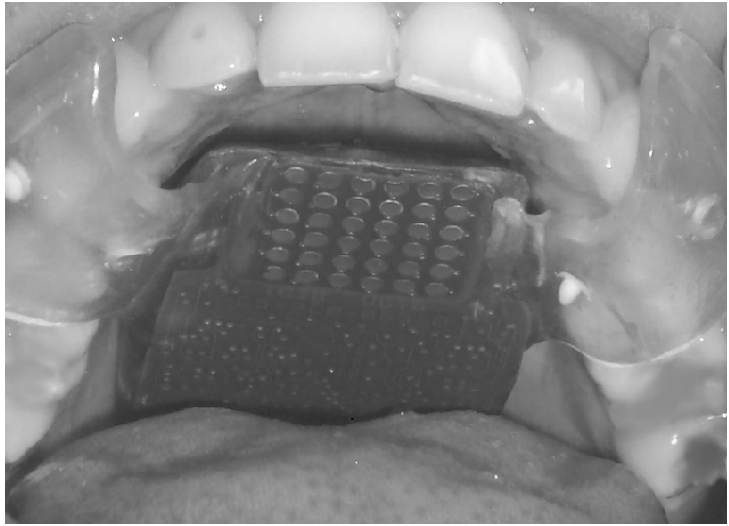

**Figure 2**

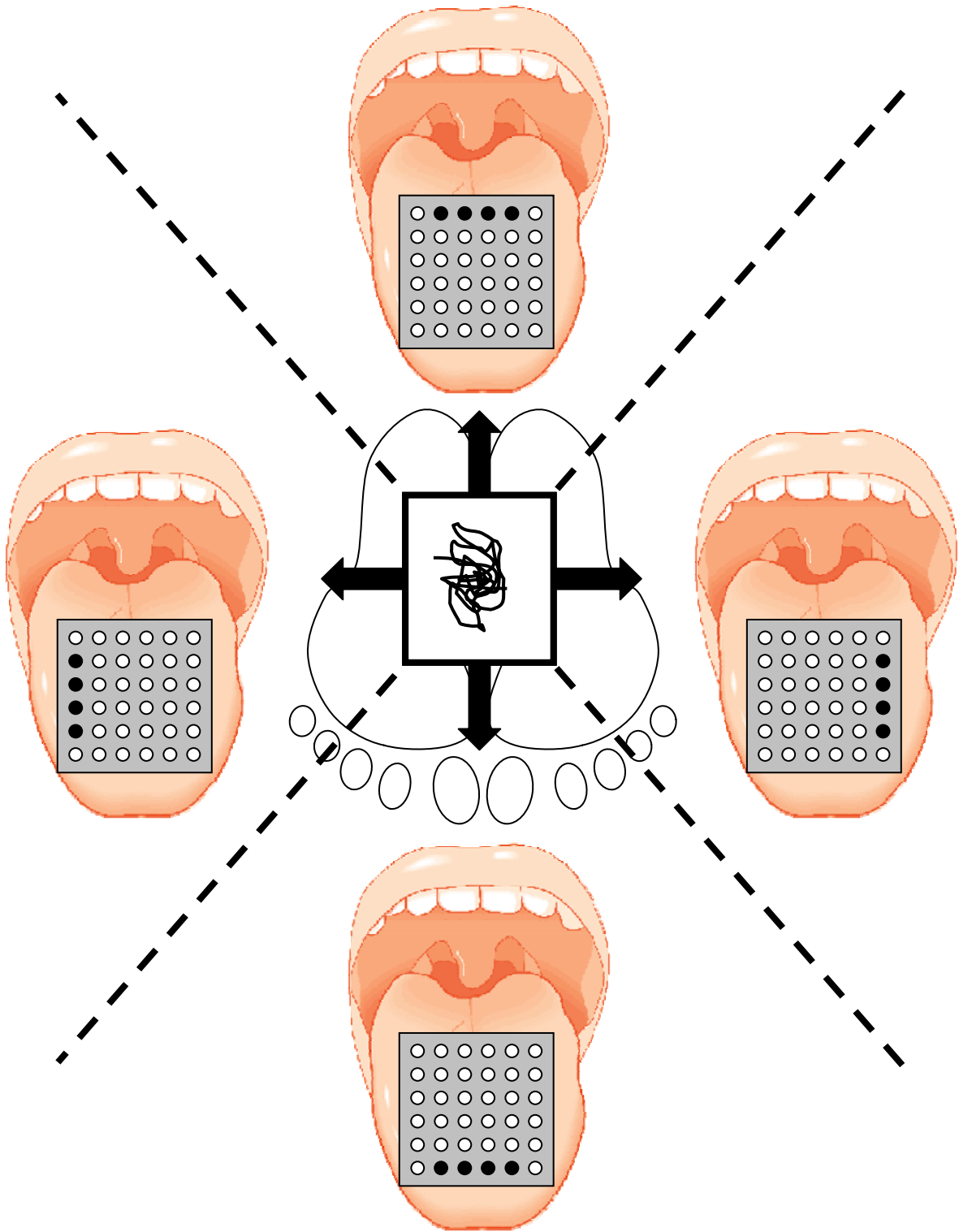

**Figure 3**

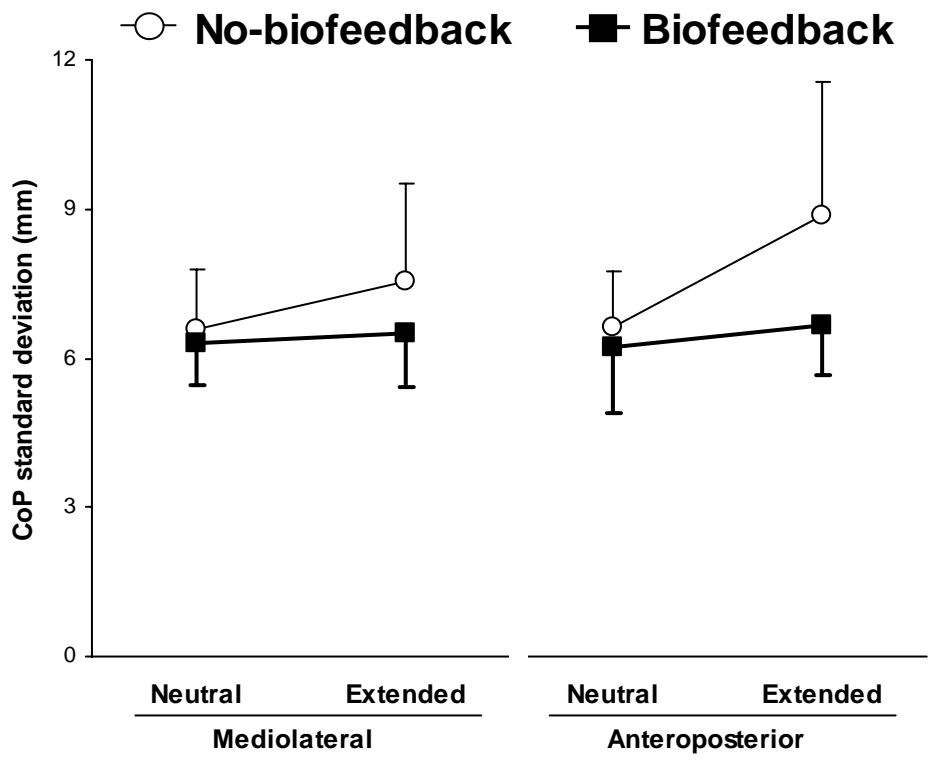

**Figure 4**

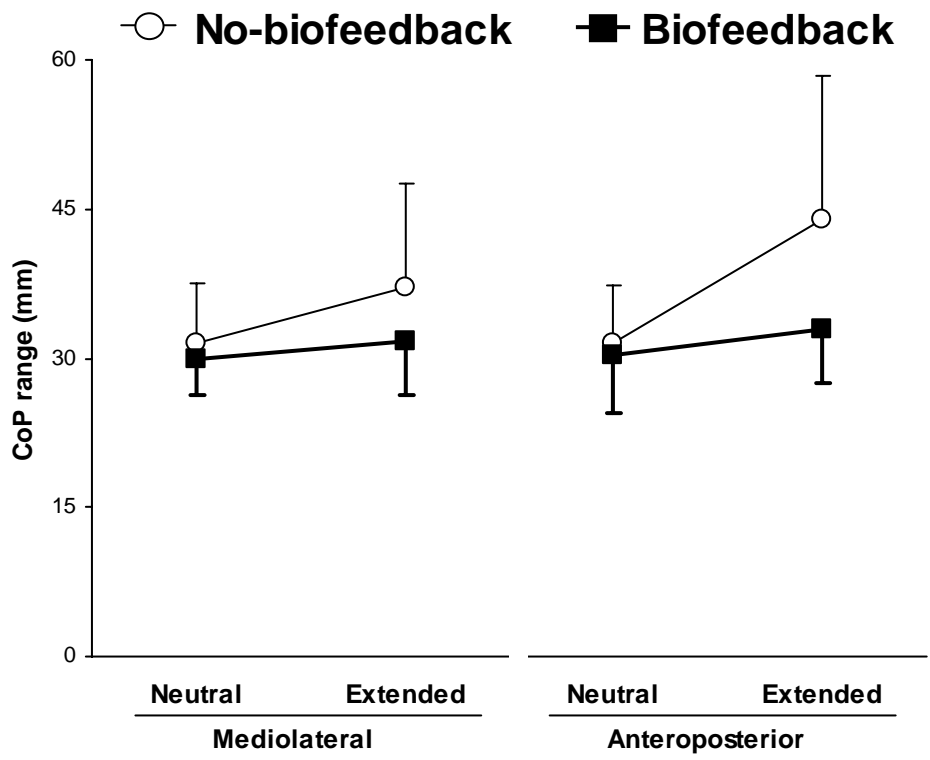